\documentclass[conference]{IEEEtran}
\IEEEoverridecommandlockouts
\usepackage{cite}
\usepackage{amsmath,amssymb,amsfonts}
\usepackage{algorithmic}
\usepackage{graphicx}
\usepackage{textcomp}
\usepackage{ifthen}
\usepackage{subfig}
\def\BibTeX{{\rm B\kern-.05em{\sc i\kern-.025em b}\kern-.08em
    T\kern-.1667em\lower.7ex\hbox{E}\kern-.125emX}}
\usepackage[linesnumbered,ruled,vlined]{algorithm2e}

\usepackage[a4paper,
            bindingoffset=0.2in,
            left=0.5in,
            right=0.5in,
            top=0.8in,
            bottom=1.2in,
            footskip=.25in]{geometry}

\begin{document}


\title{A Reduced-Complexity Maximum-Likelihood Detection with a sub-optimal BER Requirement}
\author{\IEEEauthorblockN{ Sharan Mourya}
\IEEEauthorblockA{\textit{Indian Institute of Technology, Kharagpur} \\
Hyderabad, India \\
sharanmourya7@gmail.com}\\
\and
\IEEEauthorblockN{ Amit Kumar Dutta}
\IEEEauthorblockA{\textit{Indian Institute of Technology, Kharagpur} \\
Kharagpur, India \\
amitdutta@gssst.iitkgp.ac.in}}

\maketitle

\begin{abstract}
Maximum likelihood (ML) detection is an optimal signal detection scheme, which is often difficult to implement due to its high computational complexity, especially in a multiple-input multiple-output (MIMO) scenario. In a system with $N_t$ transmit antennas employing $M$-ary modulation, the ML-MIMO detector requires $M^{N_t}$ cost function (CF) evaluations followed by a search operation for detecting the symbol with the minimum CF value. However, a practical system needs the bit-error ratio (BER) to be application-dependent which could be sub-optimal. This implies that it may not be necessary to have the minimal CF solution all the time. Rather it is desirable to search for a solution that meets the required sub-optimal BER. In this work, we propose a new detector design for a SISO/MIMO system by obtaining the relation between BER and CF which also improves the computational complexity of the ML detector for a sub-optimal BER.
\end{abstract}



\maketitle

\section{Introduction}
\label{sec:introduction}
\par 
For any modern communication system, a maximum likelihood (ML) detector is preferred due to its best data recovery performance\cite{det_est_book} but it is computationally very expensive to implement. This issue of computational complexity in conventional digital machines is problematic as it increases the latency with the increase of signal constellation size and the number of transmitter antenna (TA). Magnificent tele-traffic growth in the last decade has pushed the computational complexity and latency of a base station (BS) to an alarming level. Diversified computational requirements and massive growth in the connected user equipment (UE) in modern cellular standards make the situation worse. From the physical layer point of view, higher complexity operations include signal detection, parameter estimation, and various other error corrections at the receiver. To address this, several low complexity detection schemes were introduced in the recent years. \par 
In \cite{red1}, complexity of ML detection is reduced using Sensitive Bits (SB). \cite{red2} proposes a pre-decoder guided local exhaustive search mechanism for V-BLAST \cite{vblast}. In \cite{mpsk} a similar reduction in ML complexity is achieved in M-PSK modulated systems. Complexity reduction for a spatially modulated system  is proposed in \cite{sm}. A more generic approach with wide range of applicability is considered in \cite{orig} where performance is traded off with computational complexity. But this approach doesn't address the search algorithm, it rather deals with the complexity reduction of calculation of the cost function (CF). In this work, we try to address the search complexity aspect of ML detection.
    
In the context of ML search, the extremum values from the database of likelihood values dictate the positions of decision boundaries in the constellation space. For example, for an additive white Gaussian noise (AWGN) system, ML detector boils down to a minimum distance detector and decision boundaries are the perpendicular bisectors of the lines joining any two adjacent constellation points. Also, ML minimizes the BER for an equiprobable source. If we choose any target BER other than the optimum value, then the decision boundaries change accordingly and BER increases. We exploit this trade-off in the subsequent sections in order to reduce the search complexity in ML detection.  \par
\textit{Contributions: } 
\begin{enumerate}
    \item \rm{In} this work, we propose an ML detector which looks for a sub-optimal BER while reducing the search complexity for ML detection.
    \item The proposed detection is first applied on a single input single output (SISO) scenario and then extended to a multi-antenna case.
\end{enumerate}
\section{System Model}

\subsection{SISO}
For a SISO system, a typical data model is given as 
\begin{align}
  y = hs + w,
  \label{data_model_siso}
\end{align}
where $h,s,w$ are channel, transmitted data and AWGN, respectively. We assume that $s$ is from the constellation space $S$ with cardinality $M$. For ML detection, the likelihood function (LF) corresponding to any $s = s_i$ is given as
\begin{align}
LF(s_i) = \displaystyle \frac{1}{\sqrt{2\pi\sigma^{2}}}\displaystyle \exp\left[-\frac{1}{2}\left(\frac{y-s_i}{\sigma}\right)^{2}\right].
\label{lf}
\end{align}
ML algorithm maximizes~\eqref{lf}, which is the same as minimizing $(y-s_i)^{2}$. So we make the following search space $\Theta$ for the ML algorithm
\begin{align}
\Theta = \{d_{0}, d_{1}, ...., d_{M-1}\},
\label{theta}
\end{align}
where $d_i = \vert y - s_i\vert$ for $s_i \in S$, i.e. $d_{i}$ is the distance between the received symbol and $i^{th}$ constellation point. For ML detection, finding the optimum value is actually finding the constellation point nearest to the received symbol. Therefore, the search complexity depends on the size of $\Theta$. For a sub-optimal BER we search for a value from a desired data set other than $\Theta$. This data set, say $\Theta_d \subseteq \Theta$ contains all the values that ensure that BER is less than the maximum allowable BER. With this motivation, we first start with a binary phase-shift keying (BPSK) system.
\subsubsection{BPSK}
\begin{figure}[h]
\centering
\includegraphics[width=2 in,height=1.5 in]{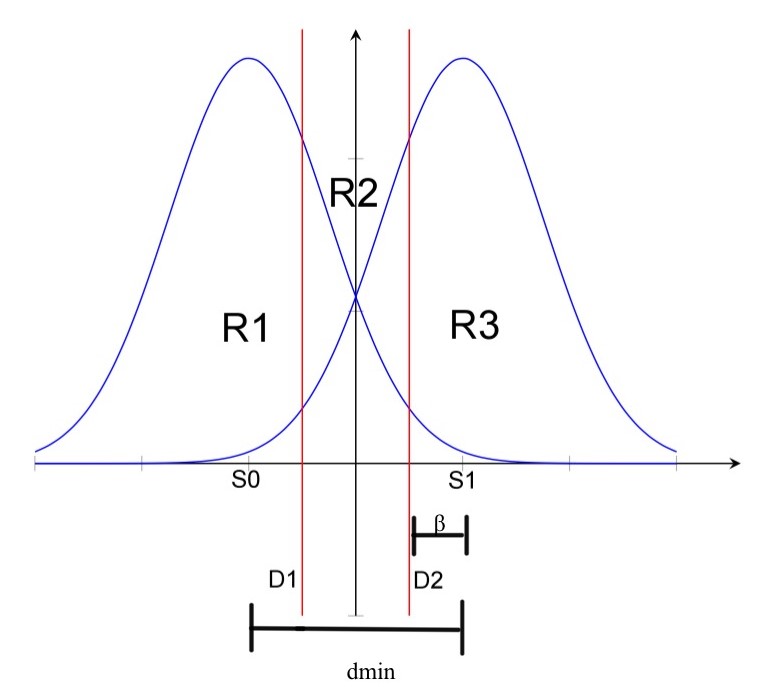}
\caption{Decision boundaries for BPSK}
\label{bpsk}
\end{figure}
Choosing the minimum value from $\Theta$ will result in decision boundaries being the perpendicular bisectors. So, when we choose a sub-optimal value, it is logical to move the decision boundary from the midpoint. This is similar to a Neyman Pearson (NP) criterion, where the boundary is shifted. However, the NP criterion is best suited for binary hypothesis, not for multiple ones. As this will lead to the determination of the decision boundary value for various constellation points and may lead to multi-objective optimization problem. To alleviate the issue, we propose to have symmetric decision regions. This leads to the concept of a "NULL" region. Hence, if the decision boundary moves in one direction by some amount, we will add a mirror image of it on the opposite side in order to preserve the symmetry. With this motivation, we propose the following detector design for BPSK. \par
Let us choose the decision boundaries to be at a distance of $\beta$ from $s_{0}$ and $s_{1}$ as shown in Fig.{~\ref{bpsk}}. Therefore, we get two decision regions ($R_{1}$ and $R_{3}$) each at a distance of $\beta$ from each constellation point. In this case, we have an overlapping region $R_{2}$ which creates ambiguity for detection. So, we propose the following detection criteria:
If $y$ is the received symbol and $\{s_{0},s_{1}\}$ form the constellation, then
\begin{enumerate}
    \item If $y$ lies in the region $R_{1}$, then $\hat{s}=s_{0}$.
    \item If $y$ lies in the region $R_{3}$, then $\hat{s}=s_{1}$.
    \item If $y$ lies in the region $R_{2}$, then decide on $\hat{s} = s_0$ or $\hat{s} = s_1$ with probability $\frac{1}{2}$.
\end{enumerate}
Let us now calculate the average probability of error for this new detector.
\begin{figure}[h]
\centering
\includegraphics[width=2.5in,height=1.5in]{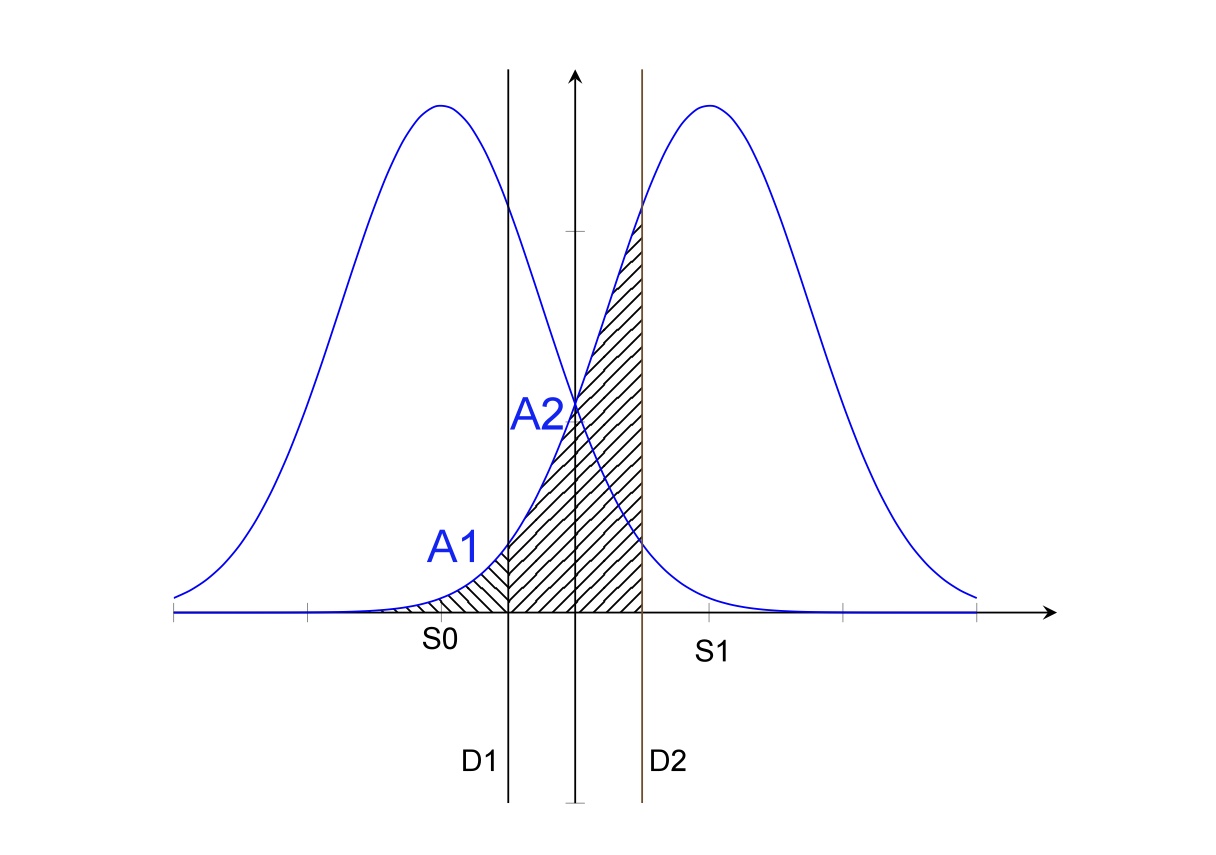}
\caption{Probability of error calculation for BPSK}
\label{decision}
\end{figure}
For $s_{1}$, error occurs when $y$ falls in the region $R_{1}$ and error might occur with probability $\frac{1}{2}$, when it falls in $R_{2}$. From Fig.{~\ref{decision}}, let us assume that $A_{1}$ is the area of the curve in $R_{1}$ and $A_{2}$ is the area of the curve in $R_{2}$. Then, the probability of error is given as
\begin{align}
P(e) & = A_{1}+\frac{1}{2}A_{2}\nonumber \\
           & = \frac{1}{2}A_{1}+\frac{1}{2}(A_{1}+A_{2}) \nonumber \\
           &  = \frac{1}{4}erfc\bigg(\frac{\vert h \vert(d_{min}-\beta)}{\sqrt{N_{0}}}\bigg) + \frac{1}{4}erfc\bigg(\frac{\vert h \vert\beta}{\sqrt{N_{0}}}\bigg).
           \label{p_bpsk}
\end{align}
For an ML detector, average probability of error is given by substituting $\beta=d_{min}/2$ as
\begin{align}
P_{min} = \frac{1}{2}erfc\bigg(\frac{\vert h \vert d_{min}}{2\sqrt{N_{0}}}\bigg).
\label{pmin}
\end{align}
We call it $P_{min}$ because it is the least value that can be achieved for any value of $\beta$ i.e., $P_e \geq P_{min}$. So by choosing a different value from $\Theta$ instead of the minimum value, we increase the probability of error. Hence, it is possible to fix a BER (application dependent) that is greater than the minimum BER and obtain the value of $\beta$ from equation \eqref{p_bpsk}. We obtain a relationship between $\beta$ and $\Theta_{d}$  in {Section}-III.B. Also, as we are addressing the issue of complexity rather than performance, we will drop the channel coefficients from hereon.

\subsubsection{PAM}
\begin{figure}[h]
\centering
\includegraphics[width=2.5in,height=1.5in]{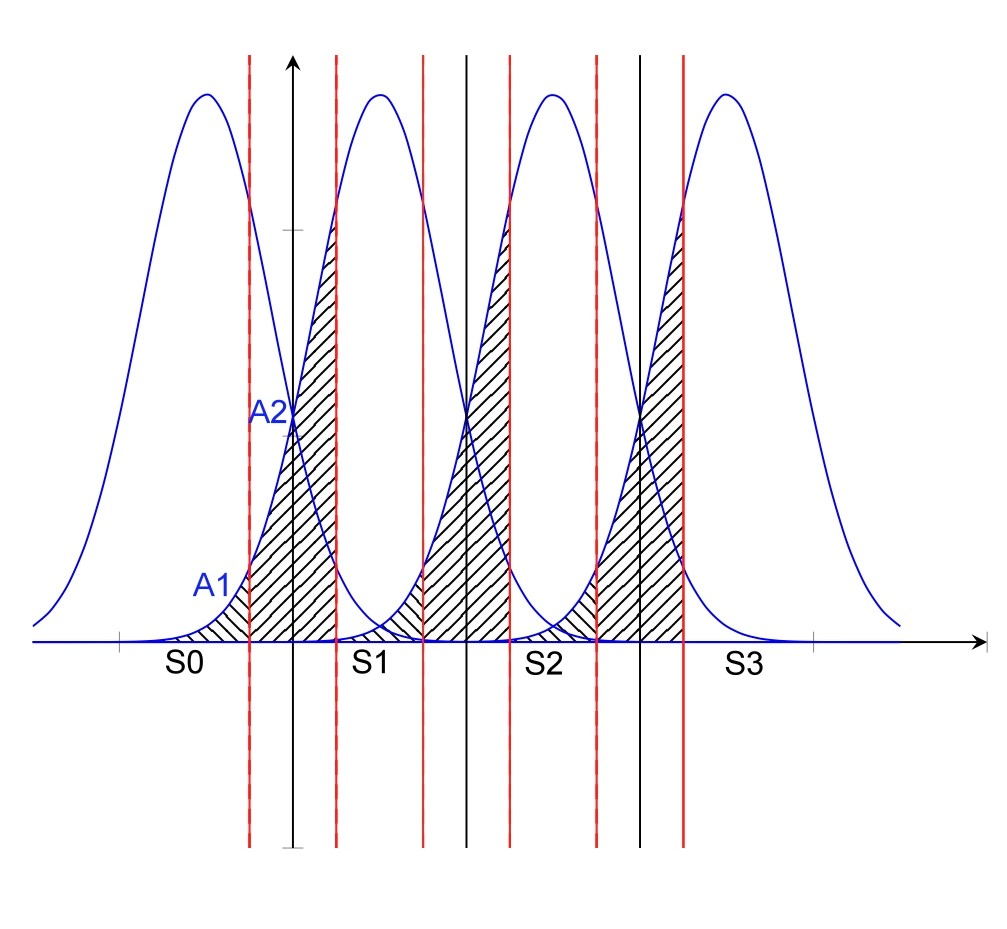}
\caption{Decision boundaries for 4-PAM}
\label{4D}
\end{figure}

Let us follow the same procedure for obtaining the average probability of error of pulse amplitude modulation (PAM). We choose 4-PAM for simplicity. The received signal distributions with the proposed decision regions are shown in Fig.{~\ref{4D}}. 
\\
\[P(e/S_{0}) = A_{1}+\frac{1}{2}A_{2} = \frac{1}{2}A_{1}+\frac{1}{2}(A_{1}+A_{2}) \]
\[P(e/S_{1}) = 2A_{1}+A_{2} = A_{1} + (A_{1} + A_{2})\]
\[P(e/S_{2}) = 2A_{1}+A_{2} = A_{1} + (A_{1} + A_{2})\]
\[P(e/S_{3}) = A_{1}+\frac{1}{2}A_{2} = \frac{1}{2}A_{1}+\frac{1}{2}(A_{1}+A_{2})\] \quad $where,$
\[A_{1} = \frac{1}{2}erfc\bigg(\frac{d_{min}-\beta}{\sqrt{N_{0}}}\bigg)\]
\[A_{1} + A_{2} = \frac{1}{2}erfc\bigg(\frac{\beta}{\sqrt{N_{0}}}\bigg)\]
\[\because P(e) = \frac{1}{4}(P(e/S_{0})+P(e/S_{1})+P(e/S_{2})+ P(e/S_{3}))\]
\[\Rightarrow P(e) = \frac{3}{4}(A_{1} + (A_{1} + A_{2}))\]
\begin{equation}
    P(e) = \frac{3}{8}erfc\bigg(\frac{d_{min}-\beta}{\sqrt{N_{0}}}\bigg)+\frac{3}{8}erfc\bigg(\frac{\beta}{\sqrt{N_{0}}}\bigg).
    \label{pam}
\end{equation}

\subsubsection{16-QAM case}
Similarly, let's derive the expression of the average probability of error for quadrature amplitude modulation (QAM). We derive it first for 16-QAM for simplicity as follows,
\begin{equation}
   P(e) = \frac{3}{4}erfc\bigg(\frac{d_{min}-\beta}{\sqrt{N_{0}}}\bigg)+\frac{3}{4}erfc\bigg(\frac{\beta}{\sqrt{N_{0}}}\bigg).
    \label{qam}
\end{equation}
The detailed derivation is given in Appendix-B.

\subsubsection{M-QAM case}
We will generalize the expression for $M$-QAM as,
\begin{align}
P(e) = \bigg(1-\frac{1}{\sqrt{M}}\bigg) \bigg[erfc\bigg(\frac{d_{min}-\beta}{\sqrt{N_{0}}}\bigg)+erfc\bigg(\frac{\beta}{\sqrt{N_{0}}}\bigg)\bigg].
\label{mqam}
\end{align}
The detail derivation is given in Appendix-C.

\subsection{UNION BOUND}
Before extending to MIMO, we derive the union bound of the proposed detector for a SISO link.

\subsubsection{SISO with M-QAM case}
To derive the union bound for the proposed symmetric based decision regions, we adopt the nearest neighbour approximation. We need to categorize the constellation depending on the number of neighbours the symbol has as per the following table.
\begin{center}
 \begin{tabular}{||c c c||}
 \hline
 Symbols & Cardinality & $\#$ of nearest neighbours \\ [0.5ex] 
 \hline\hline
 Edges & 4 & 2 \\ 
 \hline
 Interior & $(\sqrt{M}-2)^2$ & 4 \\
 \hline
 Other & $4(\sqrt{M}-2)$ & 3 \\
 [1ex] 
 \hline
\end{tabular}
\end{center}
where symbols are categorized into edges, interior points and others. Therefore, the probability of error for $M$-QAM can be bounded as
\begin{align}
P(e) \le \bigg\{ \frac{1}{2} Q\bigg(\frac{\beta}{\sqrt{2N_{0}}}\bigg) + \frac{1}{2} Q\bigg(\frac{d_{min}-\beta}{\sqrt{2N_{0}}}\bigg) \bigg \} \frac{1}{M} \sum_{i=0}^{M-1}n_{i},
\end{align}
where,
\[\frac{1}{M} \sum_{i=0}^{M-1}n_{i} = \frac{4\times 2+ 4\times (\sqrt{M}-2)^2 + 3\times 4(\sqrt{M}-2)}{M}\]
\[\Rightarrow \frac{1}{M} \sum_{i=0}^{M-1}n_{i} = 4\bigg(1-\frac{1}{\sqrt{M}}\bigg)\]
\[\Rightarrow P(e) \approx 4\bigg(1-\frac{1}{\sqrt{M}}\bigg) \bigg\{ \frac{1}{2} Q\bigg(\frac{\beta}{\sqrt{2N_{0}}}\bigg) + \frac{1}{2} Q\bigg(\frac{d_{min}-\beta}{\sqrt{2N_{0}}}\bigg) \bigg \}\]
Notice that this is exactly similar to Eq \eqref{mqam} and
it can be further simplified to
\begin{align}
P(e) \leq 4\bigg(1-\frac{1}{\sqrt{M}}\bigg) \bigg\{ \frac{1}{2} e^{-\frac{\beta^2}{4N_{0}}} + \frac{1}{2} e^{-\frac{(d_{min}-\beta)^2}{4N_{0}}} \bigg \}.
\end{align}

\subsubsection{MIMO}
Consider a $N_{r}\times N_{t}$ MIMO system (i.e., with $N_{t}$ transmitter antennas and $N_{r}$ receiver antennas) with an AWGN channel. Let $M_{i}-QAM$ be the modulation of $i^{th}$ transmitting antenna then the union bound expression is given as (from \cite{b12}),
\begin{align}
    P(e) \leq \displaystyle\frac{1}{\prod_{i=1}^{N_{t}} M_{i}} \sum_{x}\sum_{\hat{x} \neq x} P_{2}(x,\hat{x}).
\end{align}
Here, the first summation is over all possible input vectors (i.e., $\prod_{i=1}^{N_{t}} M_{i}$) and $P_{2}(x,\hat{x})$ is the pair-wise error probability given as follows
\begin{align}
P_{2}(x,\hat{x}) & = E\bigg\{Q\bigg(\frac{|x-\hat{x}|^2}{2N_{0}}\bigg)\bigg\} \nonumber \\
& = \bigg[ \frac{1}{2}(1-\mu) \bigg]^{N_{r}} \sum_{k=0}^{N_{r}-1} {N_{r}-1+k \choose k} \bigg[ \frac{1}{2}(1+\mu) \bigg]^{k},
\end{align}
where $ \mu = \displaystyle\sqrt{\frac{\bar{\gamma_{c}}}{1+\bar{\gamma_{c}}}}$ and $\bar{\gamma_{c}} = \displaystyle \frac{|x-\hat{x}|^2}{2N_{0}}$. Also, from \cite{b12}, if $\bar{\gamma_{c}} \gg 1$ then $P_{2}(x,\hat{x})$ would become
\begin{equation}
P_{2}(x,\hat{x}) = \bigg(\frac{1}{4\bar{\gamma_{c}}}\bigg)^{N_{r}} {2N_{r}-1 \choose N_{r}}.
\end{equation}
Then, the simplified union bound expression is
\begin{equation}
    P(e) \leq \frac{(2N_{0})^{N_{r}} {2N_{r}-1 \choose N_{r}}} {\prod_{i=1}^{N_{t}} M_{i}} \sum_{x}\sum_{\hat{x} \neq x} \bigg(\frac{1}{4|x-\hat{x}|^2} \bigg)^{N_{r}}.
    \label{union}
\end{equation}
\section{IMPLEMENTATION}
There are a few problems in this detector to deal with.\\
{1)} Obtaining $\beta$ from $P(e)$.\\
{2)} Translation of the proposed detector to a search algorithm.

\subsection{Proposed Solution for $\beta$}
\rm{From} Eq \eqref{qam} it is clear that it is not possible to write a closed-form expression for $\beta$ so we resort to numerical methods. We use the Newton-Raphson method to find a solution to Eq \eqref{qam}. Let $P$ be the required BER,  $P_{min}$ be the BER achieved by the ML detector and $k = log_{2}(M)$.
\begin{equation}
g(\beta) = \frac{3}{4k}erfc\bigg(\frac{d_{min}-\beta}{\sqrt{N_{0}}}\bigg)+\frac{3}{4k}erfc\bigg(\frac{\beta}{\sqrt{N_{0}}}\bigg)-P.
\end{equation}
Then, the derivative of this function is
\begin{equation}
g^{'}(\beta) = \frac{3}{2k\sqrt{\pi}}e^{-\frac{(d_{min}-x)^{2}}{N_{0}}}-\frac{3}{2k\sqrt{\pi}}e^{-\frac{x^{2}}{N_{0}}}.
\end{equation}
Using the above two equations we start the iteration with $\beta_{0}=0$ and the following update equation.
\begin{equation}
\beta_{n+1} = \beta_{n}-\frac{g(\beta_{n})}{g^{'}(\beta_{n})}.
\end{equation}



\par
\textbf{Comment:} For a MIMO system, calculation remains the same except that $g(\beta)$ becomes the Union Bound of Probability of Error we calculated in equation \eqref{union}.
\[g(\beta) = \frac{(2N_{0})^{N_{r}} {2N_{r}-1 \choose N_{r}}} {\prod_{i=1}^{N_{t}} M_{i}} \sum_{i}\sum_{j \neq i} \bigg[\bigg(\frac{1}{4|d_{ij}-\beta|^2}\bigg)^{N_{r}}\]\[+\bigg(\frac{1}{4\beta^2} \bigg)^{N_{r}}\bigg] -P.\]
\[g^{'}(\beta) = \frac{(2N_{0})^{N_{r}} {2N_{r}-1 \choose N_{r}}} {\prod_{i=1}^{N_{t}} M_{i}} 8N_{r}\sum_{i}\sum_{j \neq i} \bigg[-\beta\bigg(\frac{1}{4\beta^2} \bigg)^{N_{r}+1}\]\[+|d_{min}-\beta|\bigg(\frac{1}{4|d_{ij}-\beta|^2} \bigg)^{N_{r}+1}\bigg].\]
\subsection{Proposed simplification of Detector}
\begin{figure}[!tbp]
  \centering
  \subfloat[Decision boundaries for BPSK]{\includegraphics[width=2in,height=1.4in]{DAAD54C5-C32C-4703-9491-2ACFB8BCAAEB.jpeg} \label{BP}}
  \hfill
  \subfloat[Equivalent receiver with just one decision boundary]{\includegraphics[width=1.7in,height=1.3in]{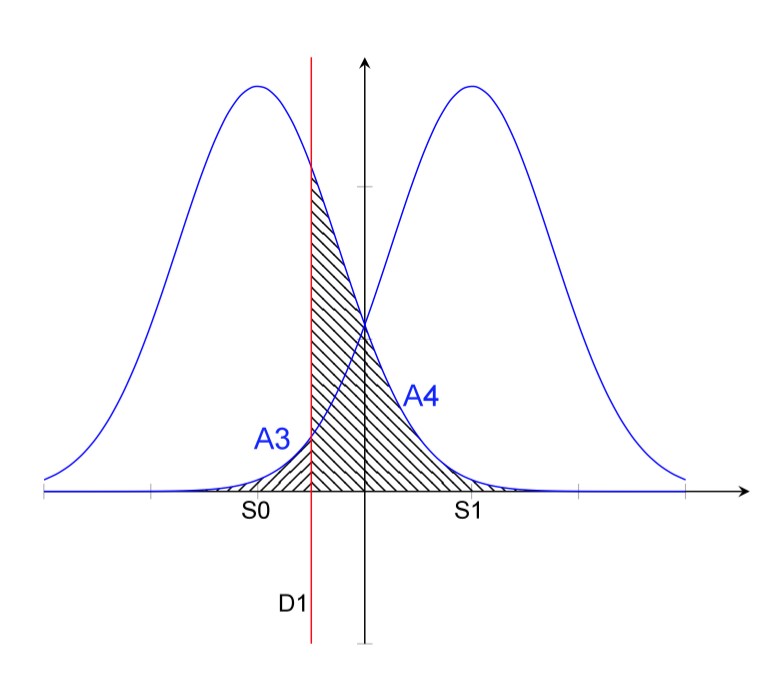}\label{NP}}
  \caption{Proposed simplification of the receiver}
\end{figure}

Implementing this detector requires a uniform distribution which adds extra circuitry to the detector. Fortunately, we can implement this detector without ever using a uniform distribution. Consider the BPSK again as in Fig.{~\ref{BP}},


Probability of error expressions for this scenario are 
\begin{align}
P_{1}(e) &= A_{1}+\frac{1}{2}A_{2} \nonumber \\
 & = \frac{1}{2}A_{1}+\frac{1}{2}(A_{1}+A_{2}).
 \end{align}
Now, consider the following scenario with only one decision boundary as in Fig.{~\ref{NP}}. Distance between $s_{0}$ and $s_{1}$ is $d_{min}$ and distance between $s_{1}$ and $D_{1}$ is $\beta$.
Probability of error in this case is given as,
\begin{align}
P_{2}(e) = \frac{1}{2}A_{3}+\frac{1}{2}A_{4}.
\end{align}
As $A_{3} = A_{1}$ and $A_{4} = A_{1}+A_{2}$, we can write 
\begin{align}
P_{2}(e) = \frac{1}{2}A_{1}+\frac{1}{2}(A_{1}+A_{2}).
\end{align}
The motivation for writing the equations in this form is clear by observing that $P_{1}(e)=P_{2}(e)$. Now, detection criteria for this detector can be written as
\begin{align}
\hat{s} & =s_{1} \quad \text{if} \quad \vert y-s_{1}\vert \leq\beta,\nonumber \\
& = s_0 \quad \text{elsewhere.}
\label{dec}
\end{align}
Given this criteria, we can now establish the relation between $\beta$ and $\Theta_{d}$. Firstly, we fix the BER at $P$ and obtain the value of $\beta$ as described in the last subsection and to practically achieve that BER we collect all the $d_{i}$'s such that $d_{i} \leq \beta$. These $d_{i}$'s form the set $\Theta_{d}$. In other words, choosing any $d_{i} \in \Theta_{d}$ from $\Theta$ will achieve the required BER. \par
\par

\section{Numerical Results}
\subsection{Probability of Hitting}
For an ML detector, probability of hitting a solution in the search algorithm for a MIMO system with $N_{t}$ transmit antennas each employed with M-QAM modulation is:
\[P_{ML}(H) = \frac{1}{M^{N_{t}}}\]
For the proposed detector, we changed the search criterion to Eq \eqref{dec} which directly reduces the search space. As we are considering an AWGN environment with mean 0 and variance $N_{0}$, all the search elements in $\Theta_{d}$ are Gaussian distributed. Thus, probability of hitting becomes:
\[P_{PD}(H) = \frac{M^{N_{t}} \times P(|y-s| \leq \beta)}{M^{N_{t}}}\]
\[P_{PD}(H) = \frac{M^{N_{t}} \times (1-Q(N_{0}\beta))}{M^{N_{t}}}\]
\[ \Rightarrow P_{PD}(H) = 1-Q(N_{0}\beta)\]
It can be see that, the proposed detector removes the exponential dependency of probability of hitting on the transmitter and constellation sizes. Now it's just a function of $\beta$ that significantly helps in scaling the approach for a massive MIMO system.
\subsection{Complexity Analysis}
To demonstrate how the proposed receiver compares to the ML receiver, we performed the following simulation under the conditions.
\\$\bullet$ \textit{Modulation:} 16-QAM with $2\times 2$ MIMO.
\\$\bullet$ \textit{Required BER:} $P(\gamma) = 2\times P_{min}(\gamma)$, where $P_{min}(\gamma)$ is the BER obtained by ML receiver at the SNR "$\gamma$".\\
We assume perfect channel estimation and equalization here.

\textbf{Procedure:}
\\\textit{1) Obtain $\beta(\gamma_{0})$ at any $\gamma_{0}$ using Newton-Raphson.}
\\\textit{2) Deploy linear search over the set $\Theta$.}
\\\textit{3) Obtain the number of evaluations the search algorithm takes to detect every symbol and average over all the symbols (10 million in this case).}
\\\textit{4) Normalize the result to 1 by dividing it with $16^2$.}
\\\textit{5) Repeat the same with the next SNR value.}
\begin{figure}[h]
\centering
\includegraphics[width=3in,height=2.5in]{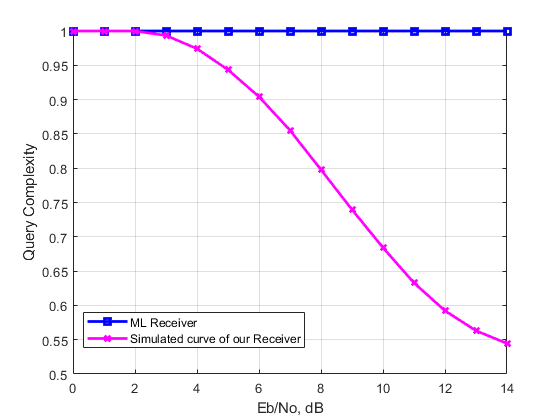}
\caption{Query Complexity vs SNR}
\label{cquery}
\end{figure} \par
ML receiver always takes the same number of evaluations, as it needs to find the least value from the set, which is equal to the size of the search space. Whereas the proposed receiver takes as many evaluations before it hits a value that is less than $\beta(\gamma)$ (i.e., a value from the set $\Theta_{d}$). This probability of hitting a solution increases with increasing SNR as the $\beta(\gamma)$ value also increases with SNR provided $P(\gamma)$ is kept constant. Hence, the number of evaluations required by the proposed receiver decreases with increasing SNR as evident from the plot in Fig.{~\ref{cquery}}.

\subsection{BER vs SNR}
\begin{figure}[]
\centering
\includegraphics[width=3in,height=2.5in]{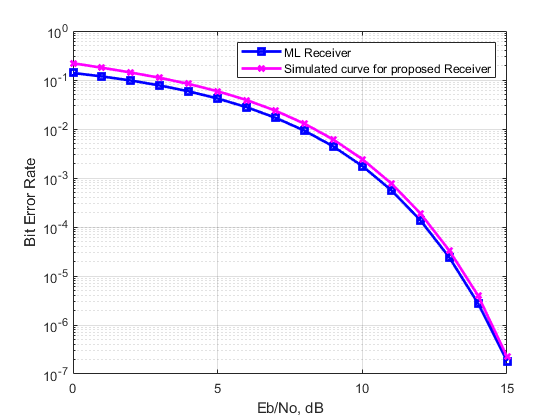}
\caption{BER vs SNR}
\label{final}
\end{figure}

We improved the computational complexity but we also do not want to trade this off with performance. For this, we simulated the BER performance of the proposed receiver on the same system as shown in Fig.{~\ref{final}}.

\section{Conclusion} In this work, we have proposed a modified ML algorithm with application dependent BER. Choosing a sub-optimal BER moves the decision boundaries such that it increases the number of solutions for the cost function which in turn reduces the size of search space for ML detection. We proposed a detector design to exploit this property while eliminating the NULL regions. Then the proposed detector is also extended to a MIMO scenario deriving the union bound expression for probability of error calculation. This detector achieved a significant computational complexity advantage depending on the target BER. Simulation results also suggests the same.  

\section*{APPENDIX}
\subsection{}
Let us consider the  detector for 4-PAM.\\

\begin{figure}[h]
\centering
\includegraphics[width=3in,height=1.4in]{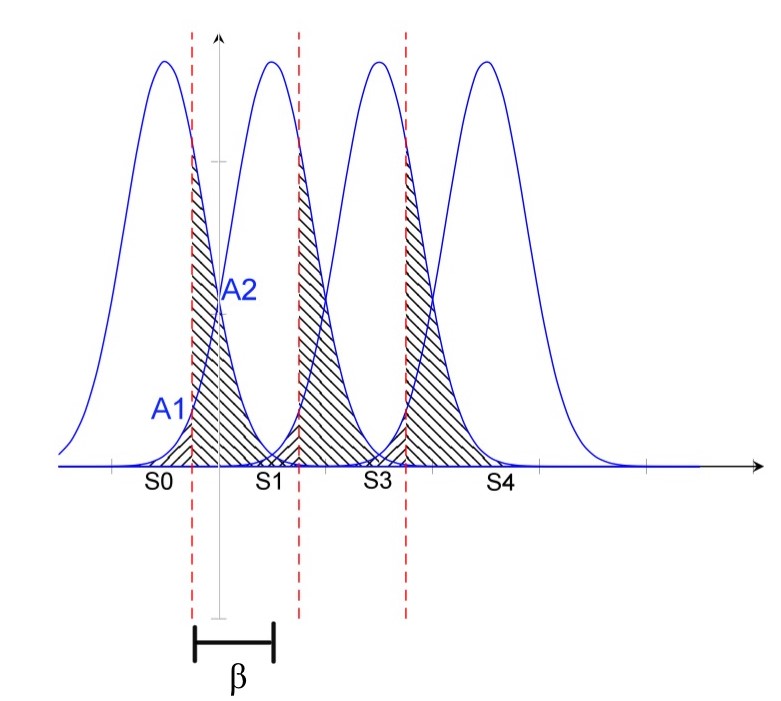}
\caption{4-PAM constellation with new boundaries}
\end{figure}
We will calculate the average probability of error for this detector and compare it with the proposed detector.
\[P(e) = \frac{1}{4}(A_{1}+(A_{1}+A_{2})+(A_{1}+A_{2})+A_{2})\]
\[P(e) = \frac{3}{4}(A_{1}+A_{2})\]
where,
\[A_{1} = \frac{1}{2}erfc\bigg(\frac{\beta}{\sqrt N_{0}}\bigg)\]
\[A_{2} = \frac{1}{2}erfc\bigg(\frac{d_{min}-\beta}{\sqrt{N_{0}}}\bigg)\]
Notice that $A_{2}$ is different from $A_{2}$ in \textit{Section II}.
\begin{equation}
    P(e) = \frac{3}{8}erfc\bigg(\frac{\beta}{\sqrt N_{0}}\bigg)+\frac{3}{8}erfc\bigg(\frac{d_{min}-\beta}{\sqrt N_{0}}\bigg).
\end{equation}

\subsection{}
Consider 16-QAM as shown in the figure. 
\begin{figure}[h]
\centering
\includegraphics[width=1.9in,height=1.7in]{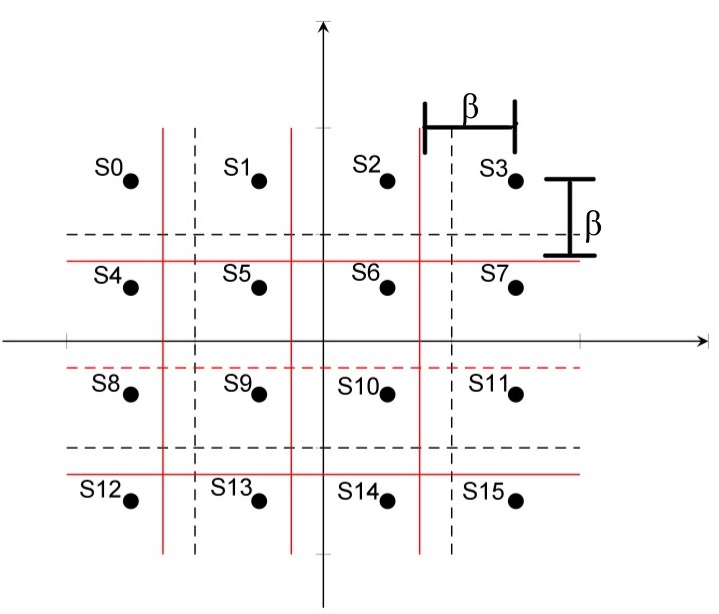}
\caption{16 QAM constellation with new boundaries}
\end{figure}
Consider the exterior points $S_{0}$, $S_{3}$, $S_{12}$, $S_{15}$. Probability of detection and probability of error of these points are given as
\[P(d/S_{0,15}) = \bigg[1-\frac{1}{2}erfc\bigg(\frac{d_{min}-\beta}{\sqrt{N_{0}}}\bigg)\bigg]\bigg[1-\frac{1}{2}erfc\bigg(\]\[\frac{\beta}{\sqrt N_{0}}\bigg)\bigg]\]
\[P(e/S_{0,15}) = 1-P(d/S_{0,15})\]
\[P(e/S_{3}) = 1-\bigg[1-\frac{1}{2}erfc\bigg(\frac{\beta}{\sqrt N_{0}}\bigg)\bigg]^{2}\]
\[P(e/S_{12}) = 1-\bigg[1-\frac{1}{2}erfc\bigg(\frac{d_{min}-\beta}{\sqrt N_{0}}\bigg)\bigg]^{2}\]
Similarly for the points  $S_{1}$,$S_{2}$,$S_{4}$,$S_{8}$,$S_{7}$,$S_{11}$,$S_{13}$,$S_{14}$ (neither interior nor exterior points) probability of detection and probability of error of these points are given by,
 \[\rightarrow P(d/S_{1,2,7,11}) = \bigg[1-\frac{1}{2}erfc\bigg(\frac{\beta}{\sqrt N_{0}}\bigg)\bigg] \times\]\[\bigg[1-\frac{1}{2}\bigg[erfc\bigg(\frac{\beta}{\sqrt N_{0}}\bigg)+erfc\bigg(\frac{d_{min}-\beta}{\sqrt{N_{0}}}\bigg)\bigg]\bigg]\]
\[P(e/S_{1,2,7,11}) = 1-P(d/S_{1,2,7,11})\]

\[\rightarrow P(d/S_{4,8,13,14}) = \bigg[1-\frac{1}{2}erfc\bigg(\frac{d_{min}-\beta}{\sqrt N_{0}}\bigg)\bigg] \times\]\[\bigg[1-\frac{1}{2}\bigg[erfc\bigg(\frac{\beta}{\sqrt N_{0}}\bigg)+erfc\bigg(\frac{d_{min}-\beta}{\sqrt{N_{0}}}\bigg)\bigg]\bigg]\]
\[P(e/S_{4,8,3,14}) = 1-P(d/S_{4,8,13,14})\]

For the interior points $S_{5}$,$S_{6}$,$S_{9}$,$S_{10}$, probability of detection and probability of error of these points are given by,
\[\rightarrow P(d/S_{5,6,9,10}) = \bigg[1-\frac{1}{2}\bigg[erfc\bigg(\frac{\beta}{\sqrt N_{0}}\bigg)+\]\[erfc\bigg(\frac{d_{min}-\beta}{\sqrt{N_{0}}}\bigg)\bigg]\bigg]^{2}\]
\[P(e/S_{5,6,9,10}) = 1-P(d/S_{5,6,9,10})\]
Total probability of error is given by,
\[P(e) = \frac{1}{16}\bigg[\big(P(e/S_{3})+P(e/S_{13})+2P(e/S_{0})\big)+\]\[\big(4P(e/S_{5})\big)+\big(4P(e/S_{1}) + 4P(e/S_{4})\big)\bigg]\]
Substituting the values and simplifying it by neglecting the second order terms gives,
\begin{equation}
    P(e) = \frac{3}{4}erfc\bigg(\frac{\beta}{\sqrt N_{0}}\bigg)+\frac{3}{4}erfc\bigg(\frac{d_{min}-\beta}{\sqrt N_{0}}\bigg)
\end{equation}

\subsection{}
We will derive the general expression for probability of error of an M-QAM. Generalizing this to M-QAM we get $4$ exterior points, $(\sqrt{M}-2)^2$ interior points and $4(\sqrt{M}-2)$ points that are neither exterior nor interior.
\[P(e) = \frac{4}{M}P(e/S_{e})+\frac{(\sqrt{M}-2)^2}{M}P(e/S_{i})+\] \[\frac{4(\sqrt{M}-2)}{M}P(e/S_{n})\]
\[\rightarrow P(e) = \frac{1}{M}\bigg[\big(P(e/S_{3})+P(e/S_{13})+2P(e/S_{0})\big)\]\[+\big((\sqrt{M}-2)^2 P(e/S_{5})\big)+\]\[\big(2(\sqrt{M}-2) P(e/S_{1}) +  2(\sqrt{M}-2)P(e/S_{4})\big)\bigg]\]
Simplification gives \\
\[P(e) = \bigg(1-\frac{1}{\sqrt{M}}\bigg) \bigg[erfc\bigg(\frac{d_{min}-\beta}{\sqrt{N_{0}}}\bigg)+erfc\bigg(\frac{\beta}{\sqrt{N_{0}}}\bigg)\bigg]\]\\\\

\bibliography{reference}
\bibliographystyle{ieeetr}

\end{document}